\documentclass[apj]{emulateapj-rtx4}

\usepackage{amssymb}
\usepackage{soul}
\usepackage{graphicx}
\usepackage{amsmath}
\usepackage[usenames]{color} 
\usepackage{lscape} 
\usepackage{rotating}
\usepackage{hyperref}

\def\gv{2002\,GV\ensuremath{_{31}}}

\begin{document}
\sloppy

\title{Pushing the limits, episode 2: K2 observations of extragalactic RR Lyrae stars in the dwarf galaxy Leo IV}
\shorttitle{Extragalactic RR Lyrae stars with K2}

\author{L.~Moln\'ar\altaffilmark{1}}
\email{molnar.laszlo@csfk.mta.hu}
\author{A.~P\'al\altaffilmark{1,2}}
\author{E.~Plachy\altaffilmark{1}}
\author{V.~Ripepi\altaffilmark{3}}
\author{M.~I.~Moretti\altaffilmark{4,5}}
\author{R.~Szab\'o\altaffilmark{1}}
%\author{Gy.~M.~Szab\'o\altaffilmark{1,3,4}}
\author{L.~L.~Kiss\altaffilmark{1,6,7}}
%\author{K.~S\'arneczky\altaffilmark{1,4}}
%\author{Cs.~Kiss\altaffilmark{1}}
\altaffiltext{1}{Konkoly Observatory, Research Centre for Astronomy and Earth Sciences, Hungarian Academy of Sciences, H-1121 Budapest, Konkoly Thege Mikl\'os \'ut 15-17, Hungary}
\altaffiltext{2}{E\"otv\"os Lor\'and Tudom\'anyegyetem, H-1117 P\'azm\'any P\'eter s\'et\'any 1/A, Budapest, Hungary}
\altaffiltext{3}{INAF -- Osservatorio Astronomico di Capodimonte, Via Moiariello 16, I-80131 Naples, Italy}
\altaffiltext{4}{INAF -- Osservatorio Astronomico di Bologna, via Ranzani 1, 40127 Bologna, Italy}
\altaffiltext{5}{Scuola Normale Superiore di Pisa, piazza dei Cavalieri 7, 56126 Pisa, Italy}
\altaffiltext{6}{Gothard-Lend\"ulet Research Team, H-9704 Szombathely, Szent Imre herceg \'ut 112, Hungary}
\altaffiltext{7}{Sydney Institute for Astronomy, School of Physics A28, University of Sydney, NSW 2006, Australia}

\begin{abstract}
We present the first observations of extragalactic pulsating stars in the K2 ecliptic survey of the $Kepler$ space telescope. Variability of all three RR Lyrae stars in the dwarf spheroidal galaxy Leo IV were successfully detected, at a brightness of $Kp \approx 21.5$~mag, from data collected during Campaign 1. We identified one modulated star and another likely Blazhko candidate with periods of $29.8\pm0.9$~d and more than 80~d, respectively. EPIC 210282473 represents the first star beyond the Magellanic Clouds for which the Blazhko period and cycle-to-cycle variations in the modulation were unambiguously measured.The photometric [Fe/H] indices of the stars agree with earlier results that Leo IV is a very metal-poor galaxy. Two out of three stars blend with brighter background galaxies in the K2 frames. We demonstrate that image subtraction can be reliably used to extract photometry from faint confused sources that will be crucial not only for the K2 mission but for future space photometric missions as well.
\end{abstract}


\section{Introduction}
\label{sec:introduction}

{\it Kepler} has provided incredible results on extrasolar planets and 
planetary systems, as well as on stellar astrophysics. The space telescope was designed to be the most precise photometer ever built in order to detect the transits of numerous small planets \citep{borucki2010}. To achieve this goal, the original mission of \textit{Kepler} focused almost exclusively on the Galactic stellar population, approximately 170~000 targets, within the Lyra-Cygnus field. But after the failure of two reaction wheels the telescope permanently lost its  ability to maintain its original attitude. Soon a new mission, called K2, was initiated to save the otherwise healthy and capable space telescope \citep{howell2014}. Since then, {\it Kepler} has been observing in shorter, 75 day long campaigns along the Ecliptic 
to balance the radiation pressure from the Sun.

The new fields opened the possibility to extend the capabilities of \textit{Kepler} into new areas and we set out to explore the limits of the K2 mission. \citet{molnar2015} investigated the first observations of field RR~Lyrae stars. \citet{szabo2015} and \citet{pal2015} showed that \textit{Kepler} can be used for Solar~System research, including the detection of main-belt asteroids and trans-neptunian objects (TNOs). In this paper we continue this work, expanding the reach of \textit{Kepler} towards extragalactic pulsating stars. The three RR~Lyrae stars in the galaxy Leo~IV are the first non-cataclysmic stellar targets \textit{Kepler} ever detected outside the Galaxy. We note in passing that the space telescope has already observed four supernovae during the original mission \citep{olling13}. 

Leo~IV is one of the small and faint dwarf spheroidal galaxies around the Milky Way that were recently discovered with the help of the Sloan Digital Sky Survey, with a mass and luminosity of $M =(1.4\pm1.5)\cdot 10^6\,\mathrm{M}_\odot$ and $M_V=-5.1\pm0.6\,$mag \citep{belokurov2007,simongeha2007}. It is located at a heliocentric distance of $154\pm5$ kpc, and has a half-light radius of 3.3~arcmin translating to a physical size of 160 pc \citep{moretti2009}. Most of the stars in Leo~IV are older than 12~Gyr, although another, brief episode of star formation likely occurred 1-2~Gyr ago \citep{sand2010}. The galaxy is very metal-poor: different studies determined the average metallicity to be around $\langle[Fe/H]\rangle = -2.3$ \citep{simongeha2007,sand2010} or possibly as low as $\langle[Fe/H]\rangle = -2.58$ \citep{kirby2008}. The galaxy was also surveyed by \citet{moretti2009} who discovered four variable stars, three fundamental-mode RR~Lyrae (RRab) and a single SX~Phe pulsator. 

RR Lyrae stars are ubiquitous in all nearby galaxies and have been detected even beyond the Local Group \citep{dacosta2010}, but our knowledge decreases with distance. Our closest neighbors, the Magellanic Clouds and the Sagittarius dwarf and its stellar stream have been surveyed extensively both in terms of stars and temporal coverage, thanks to the MACHO and OGLE programs (see, for example, \citealt{alcock1996,soszynski2009,soszynski2010,soszynski2014}). But studies of other galaxies, especially of those beyond 80-100~kpc, are based on less amounts of data. These observations are usually aimed at distance determination and population studies. 

Extragalactic surveys have their limitations, however. Multiple studies revealed several candidate Blazhko stars. \citet{stetson2014} for example, identified 24 stars with signs of amplitude variation out of 194 RR~Lyrae stars  in Leo I, and we know one candidate in M31 too \citep{brown2004}. These low occurrences are likely just lower limits: we now know that about $50\%$ of RRab stars are modulated in the Milky Way \citep{jurcsik2009,benko2014}. But in order to determine the accurate occurrence rates and modulation periods of Blazhko stars, one usually requires several weeks of intensive observations, and thus unambiguous detections have not yet been made. Also, data from $Kepler$ revealed that continuous observations are the key to detect any cycle-to-cycle variations---another feature that earlier studies lacked \citep{szabo2010}. Therefore we set out to gather the first extended, uninterrupted observations from extragalactic pulsating stars with the help of K2. 

%% %% %% %% %% %% %% %% %% %% %% %% %% %% %% %% %% %% %% %% %% %% %% %% %% %% 
\begin{figure}
\includegraphics[width=1.0\columnwidth]{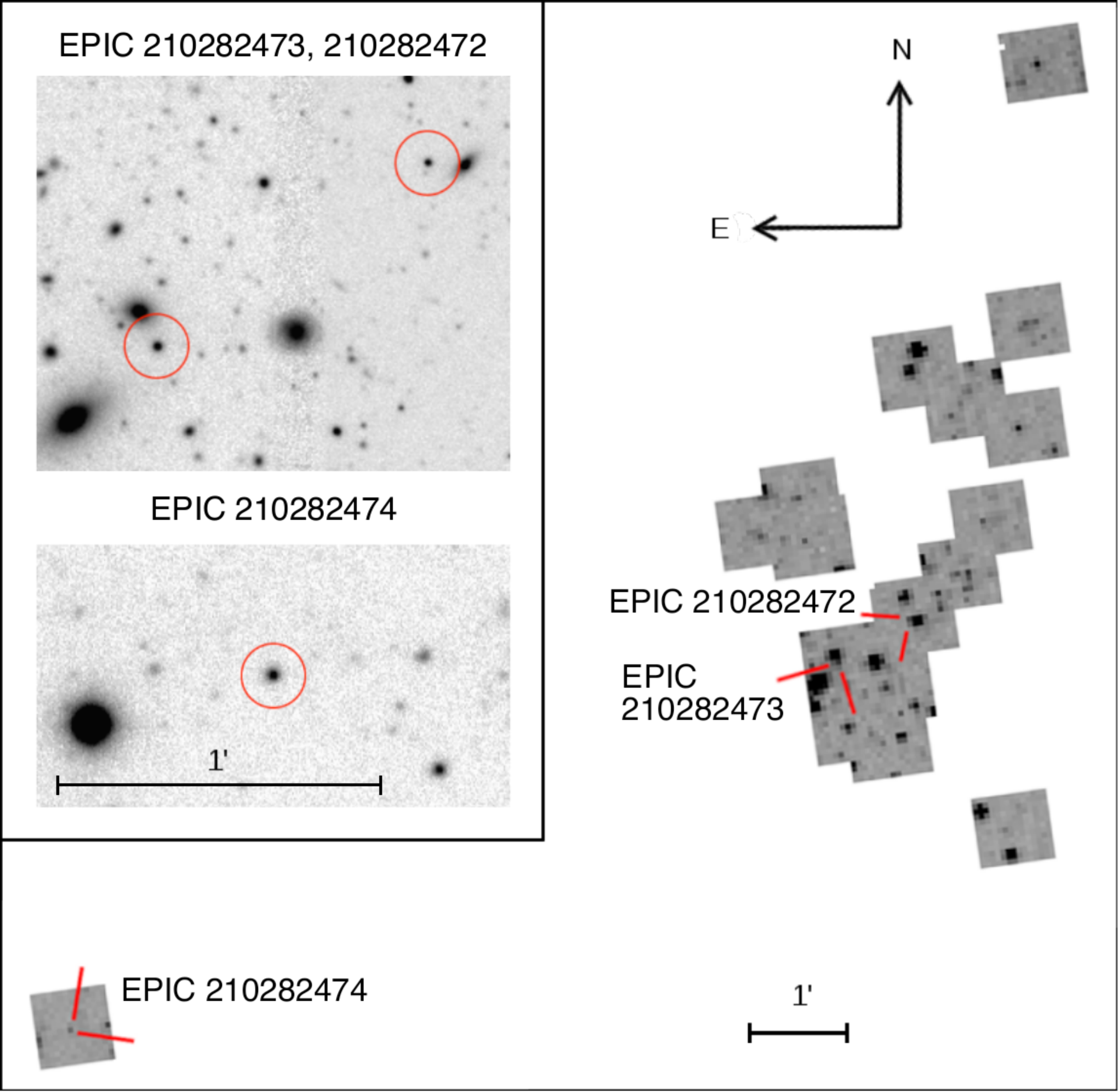}
\caption{The K2 target pixel masks covering parts of Leo IV. The three RR Lyrae stars are indicated. The insert shows high-resolution deep imagery obtained by combining data from the William Herschel Telescope, the Isaac Newton Telescope, and the Southern Astrophysical Research telescope, collected by \citet{moretti2009}. Red circles represent the apertures used in the reduction of the K2 data (see Figure \ref{fig:stamps}).}\label{fig:leo4_map}
\end{figure}
%% %% %% %% %% %% %% %% %% %% %% %% %% %% %% %% %% %% %% %% %% %% %% %% %% %%

\section{Observations and data reduction}
\label{sec:obsdatareduction}

Leo IV fell into the field-of-view of the first scientific campaign (C1) of the K2 mission and was observed between May 30 and August 21, 2014, corresponding to Barycentric Julian Day (BJD) 2456808.18--2456890.33. We requested to observe the 3 RR~Lyrae variables and the 16 brightest giant stars in the galaxy through the K2 Guest Observer proposal GO1019 in long cadence mode, with an integration time of 29.4 min. We estimated the maximum brightness of the RR~Lyrae stars to be around $Kp=21-21.2$ mag in the $Kepler$ passband, leading to a precision of a few tenths of a magnitude per long cadence data point.  At that time the much fainter SX~Phe star ($V=23.0$ mag) seemed to be beyond the capabilities of \textit{Kepler} so we did not include it in the proposal, although the recent experiences with the similarly faint TNO, \gv, suggest that we could have detected its variation too \citep{pal2015}. In this paper we focus on the three RR~Lyrae stars from the sample. The summary of their observation parameters is shown in Table~\ref{table:obs}.

%% %% %% %% %% %% %% %% %% %% %% %% %% %% %% %% %% %% %% %% %% %% %% %% %% %% 
\begin{deluxetable}{lccccr}
\tabletypesize{\scriptsize}
\tablecolumns{6}
\tablewidth{0pc}
\tablecaption{Summary of the K2 observations of the RR Ly targets in
the dwarf galaxy Leo IV.\label{table:obs}}
\tablehead{Object & Mask & R.A. & Decl. & $V$ & ID \\
 ~ & (px) & (J2000) & (J2000) & (mag) & ~ }
\startdata
 210282472 & 15$\times$14 
		& $11^{\rm h}32^{\rm m}55.8^{\rm s}$ 
		& $-0^\circ33^{\prime}29.4^{\prime\prime}$ & 21.46 & V2 \\
210282473 & 15$\times$13 
		& $11^{\rm h}32^{\rm m}59.2^{\rm s}$ 
		& $-0^\circ34^{\prime}03.6^{\prime\prime}$ & 21.47 & V1 \\
210282474 & 14$\times$14 
		& $11^{\rm h}33^{\rm m}36.6^{\rm s}$ 
		& $-0^\circ38^{\prime}43.3^{\prime\prime}$ & 21.52 & V3
\enddata
\tablecomments{IDs refer to the identifications given by \citet{moretti2009}.}
\end{deluxetable}
%% %% %% %% %% %% %% %% %% %% %% %% %% %% %% %% %% %% %% %% %% %% %% %% %% %% 

\begin{figure}
\begin{center}
\begin{tabular}{ccc}
210282472 & 210282473 & 210282474 \\[3mm]
\resizebox{25mm}{!}{\includegraphics{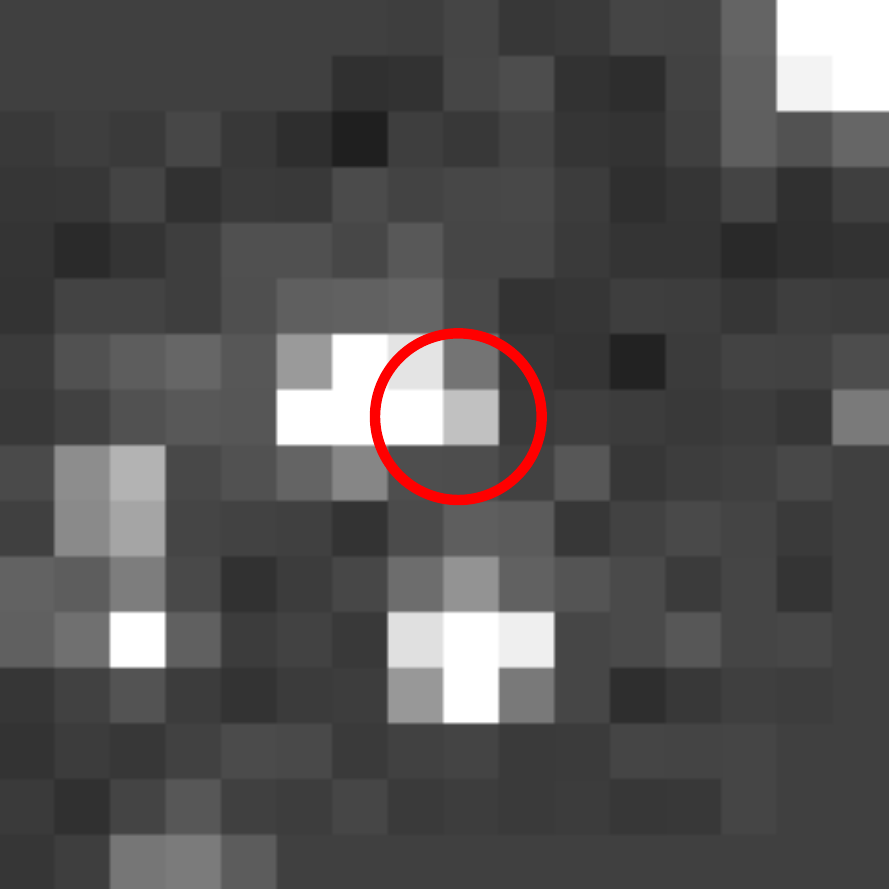}} & 
\resizebox{25mm}{!}{\includegraphics{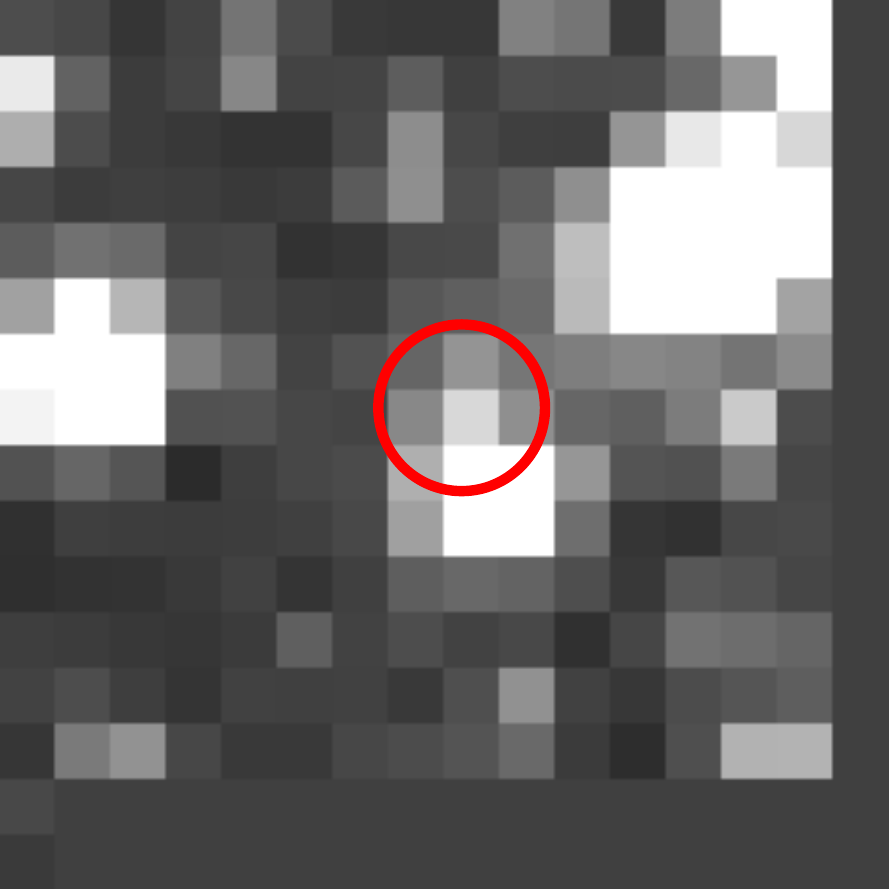}} & 
\resizebox{25mm}{!}{\includegraphics{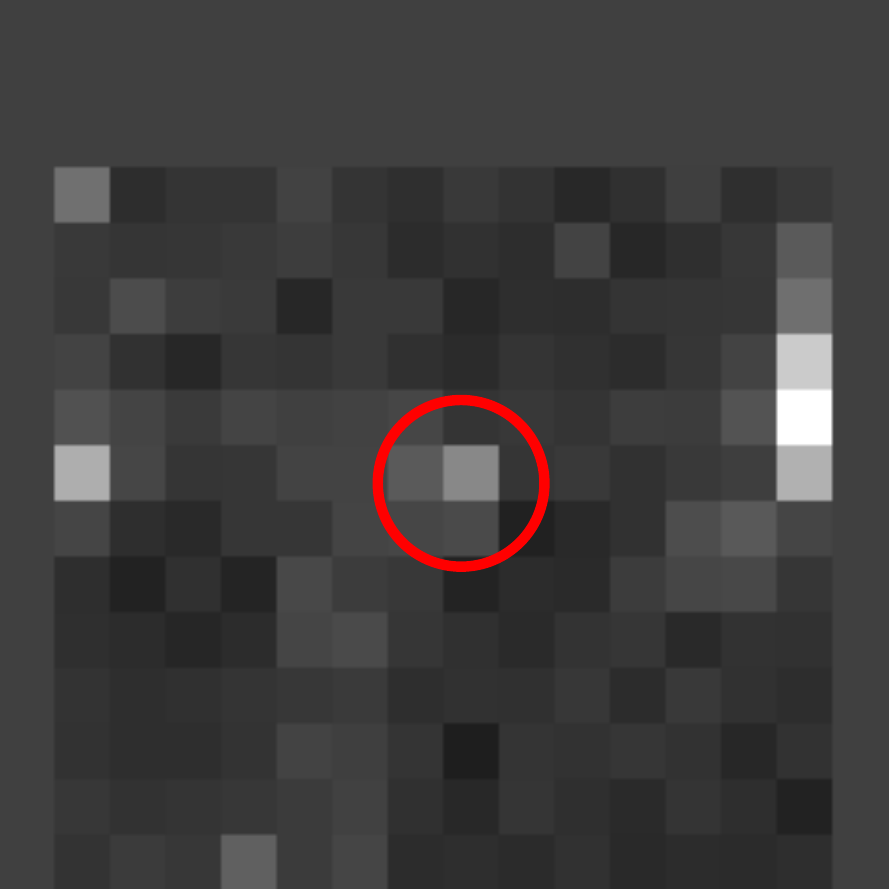}} \\[2mm]
\resizebox{25mm}{!}{\includegraphics{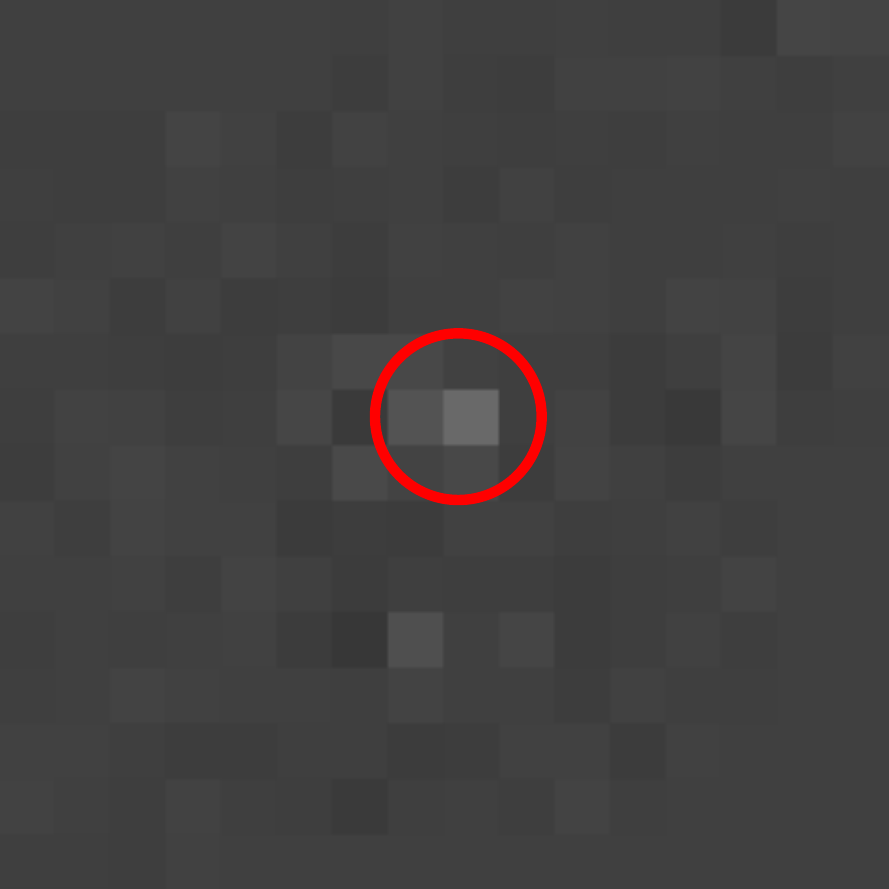}} & 
\resizebox{25mm}{!}{\includegraphics{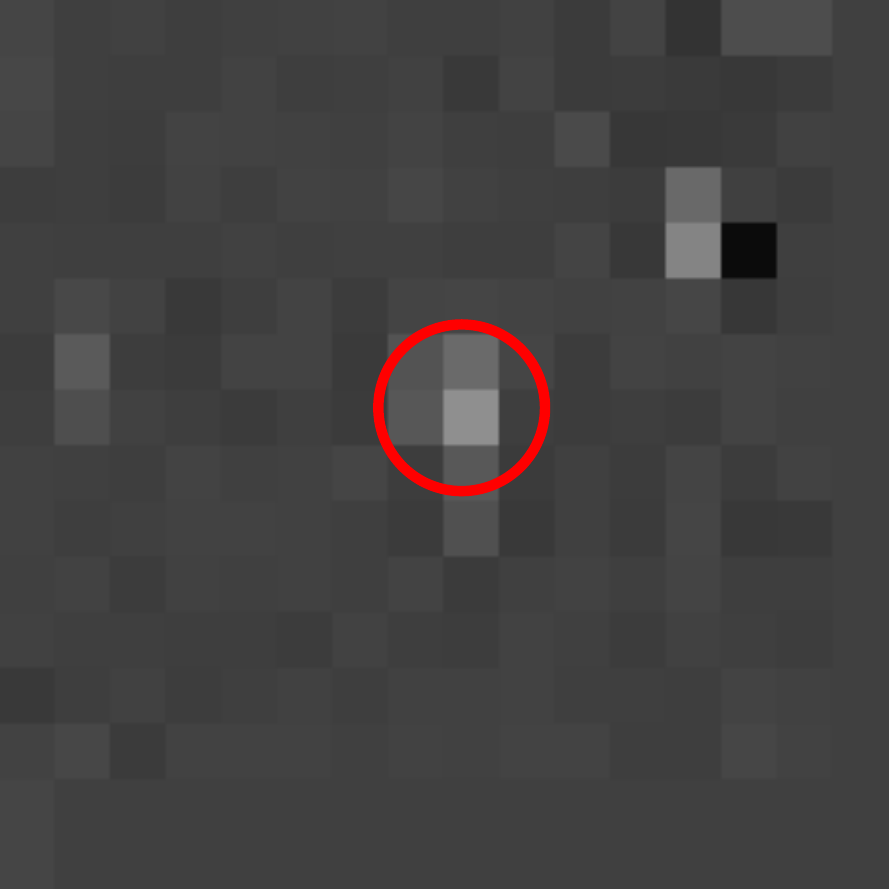}} & 
\resizebox{25mm}{!}{\includegraphics{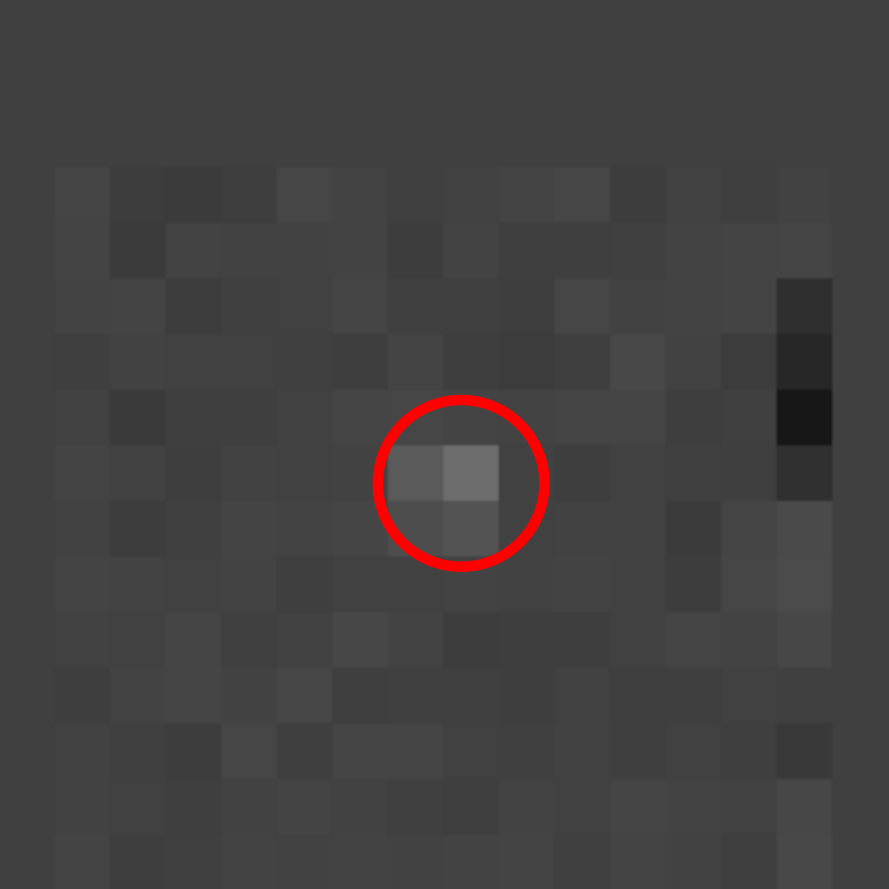}} \\[2mm]
\resizebox{25mm}{!}{\includegraphics{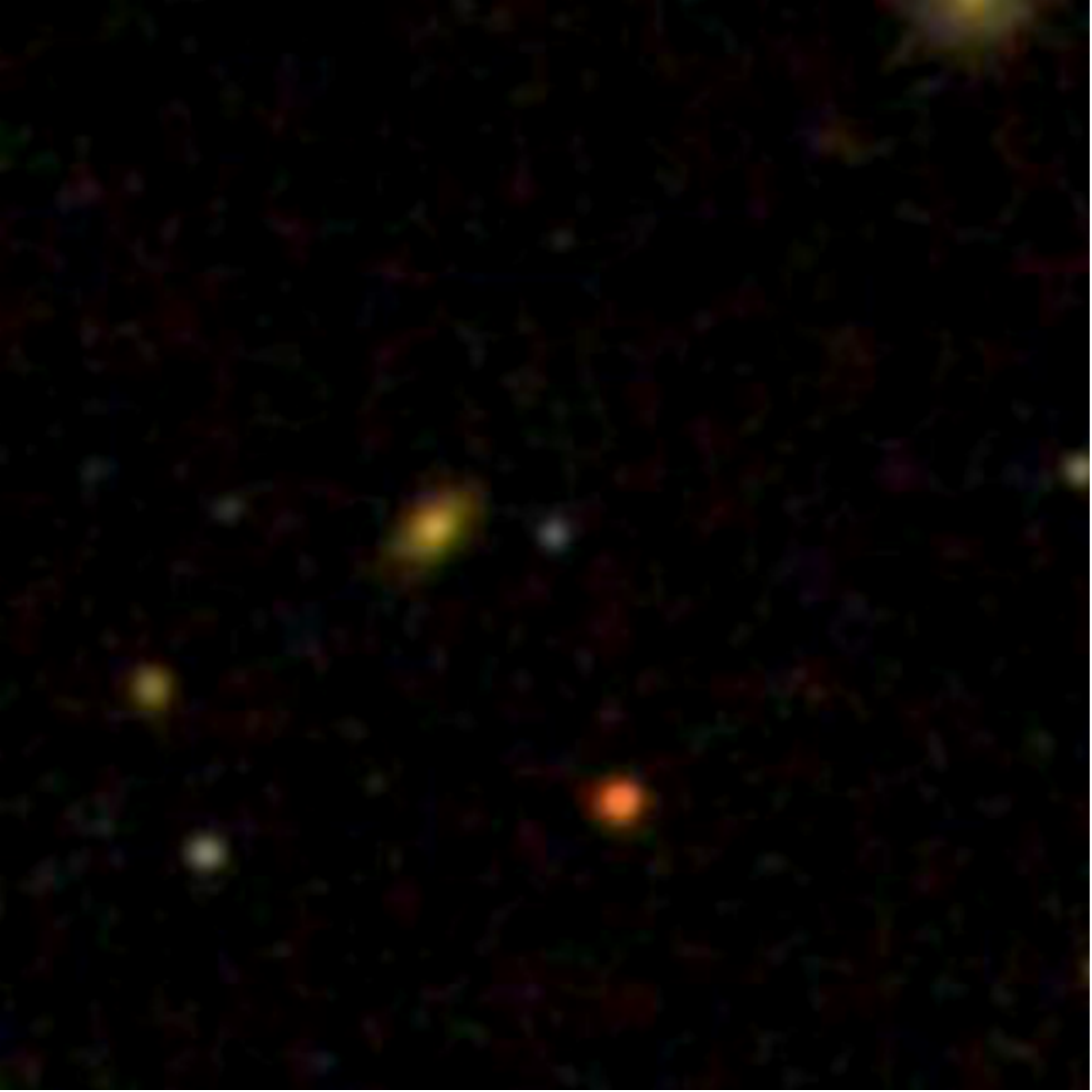}} & 
\resizebox{25mm}{!}{\includegraphics{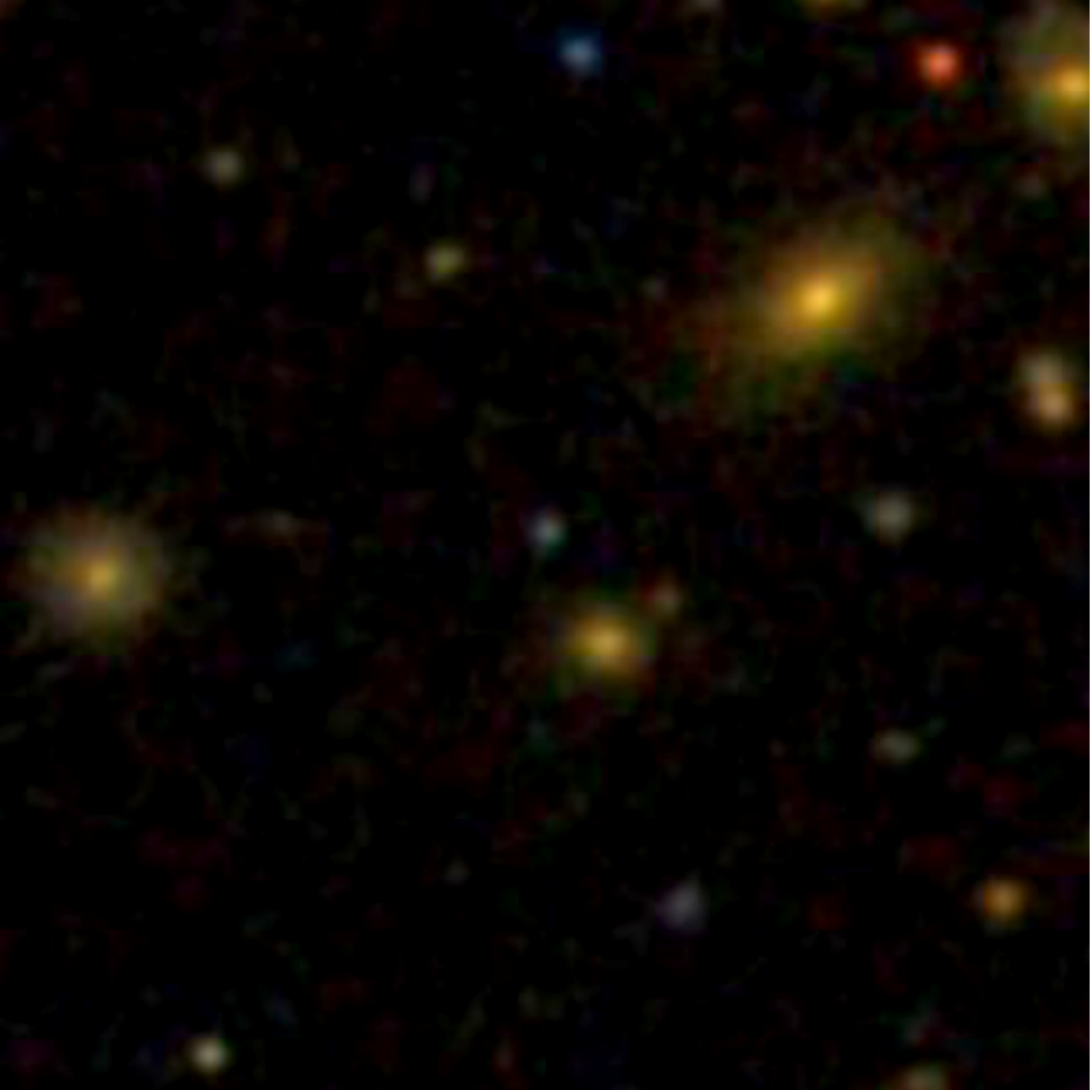}} & 
\resizebox{25mm}{!}{\includegraphics{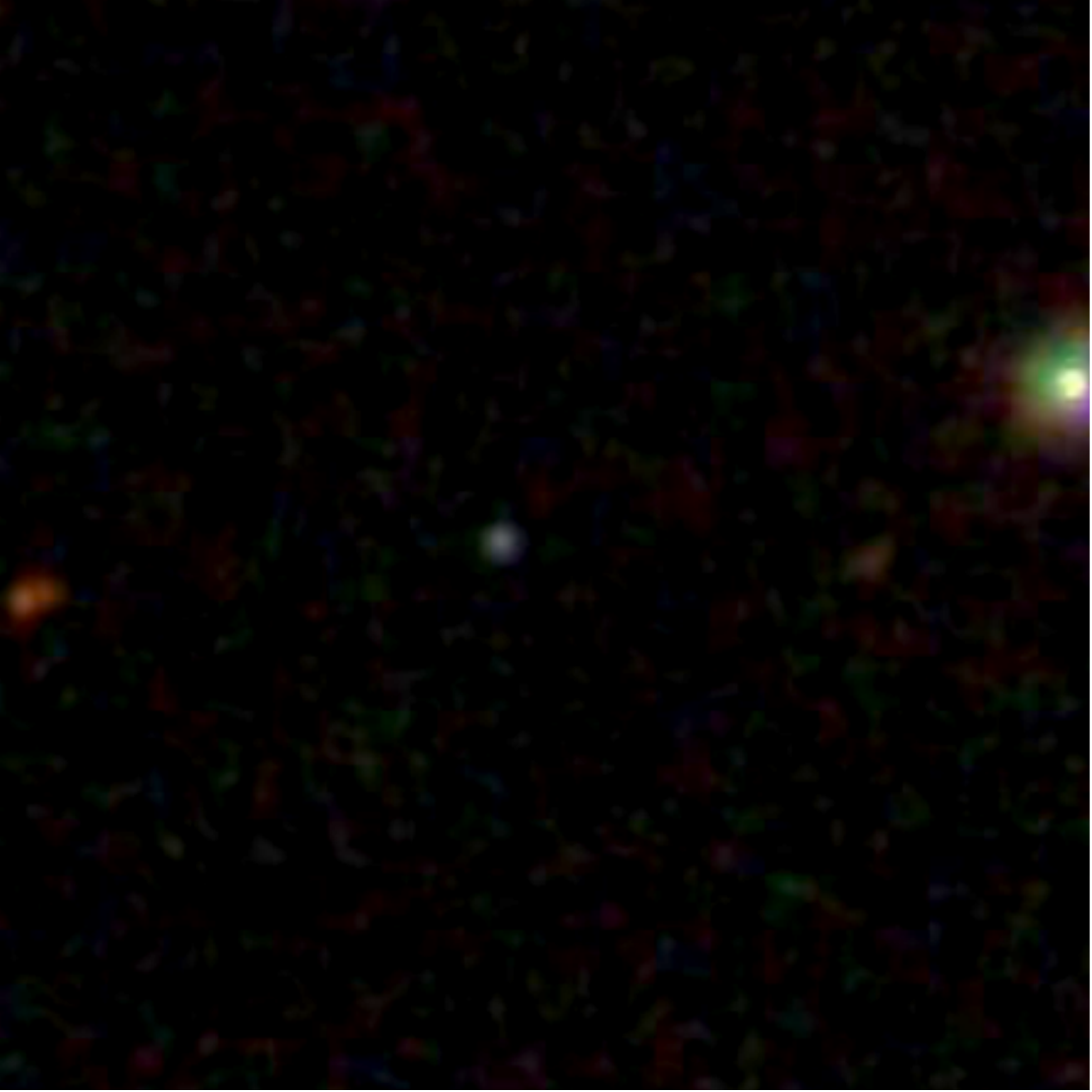}} 
\end{tabular}
\end{center}
\caption{Image stamps showing the vicinity of the K2 extragalactic RR~Lyrae
targets 210282472, 210282473 and 210282474. The first row shows the stamps
on the master median-combined image used in the process of differential photometry.
The second row shows stamps created as the average of a roughly dozen differential
images taken at the brightest phase of the RR Lyr oscillations. The 
photometric aperture with an $r=1.5$~px radius used in the procedure 
is shown as a red circle. Stamps in the third row show the respective
SDSS DR9 images. All of these stamps cover an area of 
$64^{\prime\prime}\times64^{\prime\prime}$ on the sky, equivalent to
$16\times16$ {\it Kepler} pixels.}\label{fig:stamps}
\end{figure}

For all of the tasks described below we employed exclusively the utilities 
shipped within the \textsc{Fitsh}\footnote{\href{http://fitsh.szofi.net/}{http://fitsh.szofi.net/}} \citep{pal2012}. \textsc{Fitsh} is a lightweight, yet comprehensive, 
fully open source astronomical data reduction and analysis software package. 
It is comprised of a collection of standalone binary programs that are utilized through various UNIX shell scripts. 
The accurate photometric time series for these three extragalactic Leo~IV 
RR~Lyr targets have been obtained as follows. The image scale of \textit{Kepler} 
is 3.98$^{\prime\prime}$/px, and the point spread functions are at least 3--5~px 
wide that leads to confusion of nearby sources. Although Leo IV is sparse enough to be
resolved even with $Kepler$, two out of the three RR Lyr
targets were extremely close to bright background galaxies (i.e., nearly 
within a pixel, see Figures \ref{fig:leo4_map} and \ref{fig:stamps}), 
accurate photometry can only be done by involving image 
subtraction techniques. Since the pointing jitter of {\it Kepler} in 
the K2 mission was in the range of a pixel, the individual frames 
had to be adjusted to the same reference system before performing
any kind of differential image analysis. In order to precisely estimate
the shifts and rotations between the (subsequent) frames, we involved 
additional K2 frames since the raw K2 data do not include a frame-by-frame centroid coordinate. 
These additional $14$ fields were the target pixel files of EPIC~201424914, 201430029 
and the $12$ stamps having an identifier between 210282475 and 210282486,
all from the GO1019 proposal (where EPIC refers to the K2 Ecliptic Plane Input Catalog). 
These $3+14$ stamps were combined to a 
single image and the foreground stars were used to obtain precisely the 
transformation between these images (Figure \ref{fig:leo4_map}). 

%% %% %% %% %% %% %% %% %% %% %% %% %% %% %% %% %% %% %% %% %% %% %% %% %% %% 
%% %% %% %% %% %% %% %% %% %% %% %% %% %% %% %% %% %% %% %% %% %% %% %% %% %% 

Since the K2 mission observes along the ecliptic plane in an approximately
10$^\circ$-wide area, the background of these observations also varies 
gradually due to the increasing amount of zodiacal light throughout the 
campaign. Hence, prior any differential analysis, the background must also 
be subtracted. Luckily, the fields containing the $3$ RR~Lyr targets as 
well as the additional $14$ stamps were not crowded, and the background can simply be 
determined by considering the median of all observed pixel values. 

After the frames were registered to the same reference system and
the background was subtracted accordingly, we chose every 20th image to 
obtain median-averaged master frame used as a reference for image subtraction.
Since the number of stars having a good signal-to-noise ratio in these frames were 
insufficient and the instrument does not show any significant variation in 
its point-spread function, we did not perform any cross-convolution
between the adjusted images. This procedure also simplified the evaluation
of the photometry and hence it was not necessary to perform photometry on
convolved apertures \citep{pal2009}. Therefore, the final fluxes were
obtained as the sum of the flux in the subtracted frame and in the master frame, i.e.,
$F_i = F_{i,subtr}+F_{master}$.
For these flux estimations we used simple aperture photometry with an aperture radius of $r=1.5$~px 
(Figure \ref{fig:stamps}). However, we must note that the reference
flux on the master frame can only be estimated with an additional
systematic uncertainty in the case of the confused sources (210282472 
and 210282473). Namely, we compared the peak pixel values on the stellar
sources with the fluxes obtained using an aperture of $r=1.5$~px.
The ratio of these two numbers aid the estimation of the reference flux
in the case of the confused sources as well and we found that the systematic 
uncertainty of this method is in the range of 2-3\% for these sources.
Due to the nature of the differential photometry, this kind of systematics
yields a systematic amplification in the variability amplitudes.
The random errors in the photometric fluxes have been estimated using 
the photon noise derived from the instrumental gain of
$113.3\,{\rm e^-/ADU}$ as well as the standard deviation of the background
pixels near the target sources. Due to the differential processing, this latter
component also includes the photon noise of the nearby confusing sources. 

%% %% %% %% %% %% %% %% %% %% %% %% %% %% %% %% %% %% %% %% %% %% %% %% %% %% 
\begin{figure*}
\includegraphics[width=1.0\textwidth]{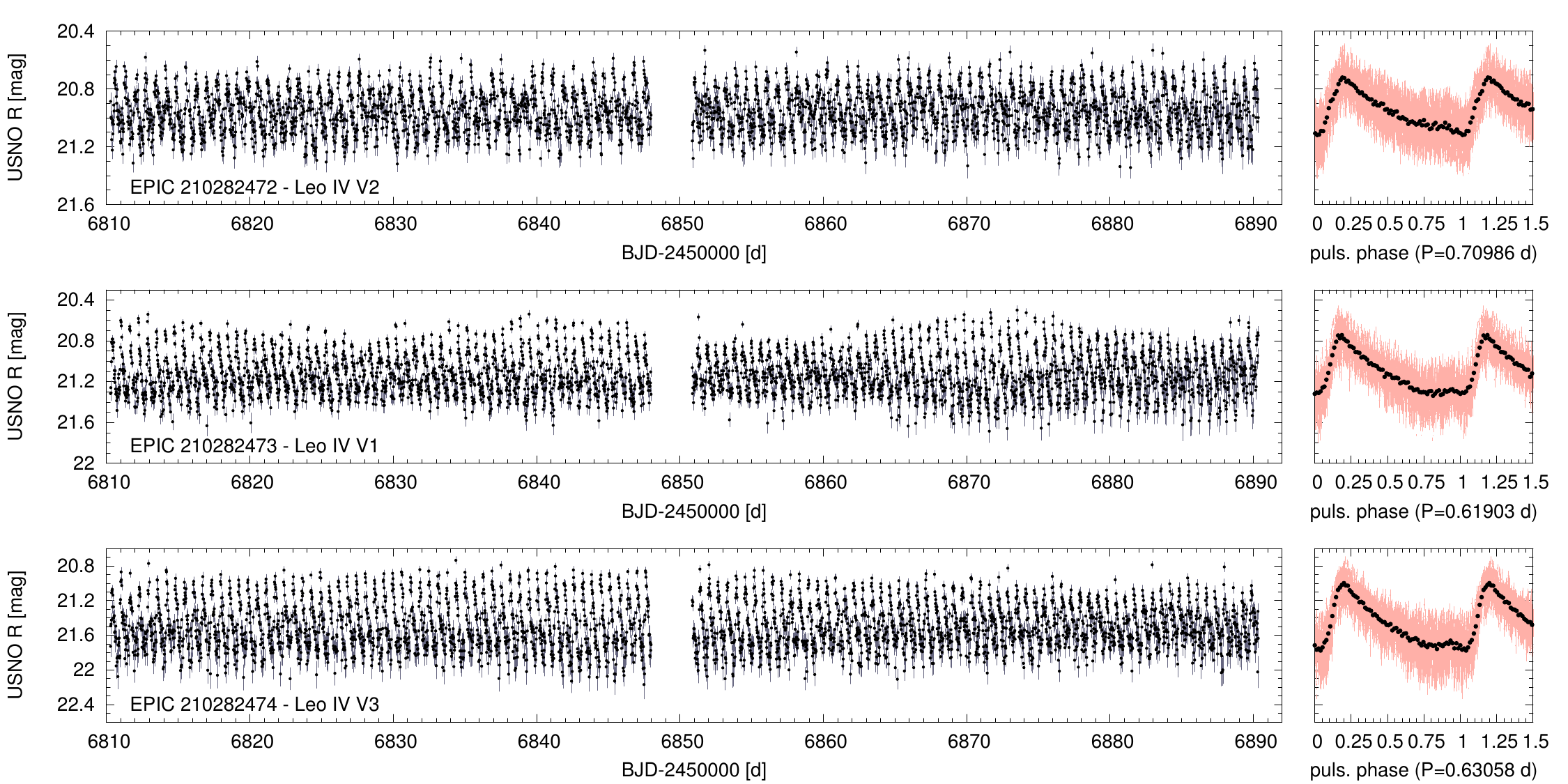}
\caption{Light curves of the three RR Lyrae stars. Left panels: light curves after outlier removal and Fourier filtering. Note the variable amplitude of EPIC~210282473 in the middle panel. Designations from \citet{moretti2009} are also indicated. The right panels show the folded light curves (pink dots and lines) and the binned phase curves (black points). We used 75 bins per pulsation period for each star.}\label{fig:lcs}
\end{figure*}
%% %% %% %% %% %% %% %% %% %% %% %% %% %% %% %% %% %% %% %% %% %% %% %% %% %%

%% %% %% %% %% %% %% %% %% %% %% %% %% %% %% %% %% %% %% %% %% %% %% %% %% %% 
\begin{deluxetable}{lcccr}
%\tabletypesize{\scriptsize}
\tablecolumns{5}
\tablewidth{0pc}
\tablecaption{K2 photometric data of the RR Lyr stars in Leo IV.\label{table:phot}}
\tablehead{Object & \colhead{Time} & \colhead{Brightness} & \colhead{Error} & Type \\ 
~ & \colhead{(BJD)} &  \colhead{($R$ mag)\ensuremath{^{\rm a}}} &  \colhead{(mag)} & ~}

\startdata
210282472	&	2456810.28292 & 20.865 &  0.062 & raw \\
210282472	&	2456810.30336 & 20.659 &  0.048 & raw \\
210282472	&	2456810.32379 & 20.974 &  0.058 & raw \\
\multicolumn{5}{l}{\ldots}\\
\hline
210282473	&	2456810.28292 & 21.247 &  0.069 & raw \\
210282473	&	2456810.30336 & 21.188 &  0.069 & raw \\
210282473	&	2456810.32379 & 21.240 &  0.076 & raw \\
\multicolumn{5}{l}{\ldots}\\
\hline
210282474	&	2456810.28292 & 21.821 &  0.122 & raw \\
210282474	&	2456810.30336 & 21.585 &  0.093 & raw \\
210282474	&	2456810.32379 & 21.565 &  0.085 & raw \\
\multicolumn{5}{l}{\ldots}\\
\hline 
210282472 & 2456810.28292 & 20.886 & 0.062 & proc \\
210282472 & 2456810.32379 & 20.995 & 0.058 & proc \\
210282472 & 2456810.34422 & 21.004 & 0.054 & proc \\
\multicolumn{5}{l}{\ldots}
\enddata
\tablecomments{Table \ref{table:phot} is published in its entirety in the
electronic edition of the {\it Astrophysical Journal Letters}.  A portion is
shown here for guidance regarding its form and content.}
\tablenotetext{a}{Magnitudes shown here are transformed to USNO-B1.0 $R$ system, see text for further details. Indices ``raw" and ``proc" correspond to raw and processed (sigma-clipped and Fourier-filtered) data, respectively. }
\end{deluxetable}
%% %% %% %% %% %% %% %% %% %% %% %% %% %% %% %% %% %% %% %% %% %% %% %% %% %% 

\newpage

\section{Light curve analysis}
\label{sec:analysis}

Once the raw light curves were obtained, we removed the outliers from the light curves with a sigma-clipping algorithm. We subtracted an initial Fourier fit from the light curves, clipped the residual at the $3\sigma$ levels three consecutive times and then restored the original signal. 

Finally, we applied Fourier filtering to remove the slow variations. The photometry of these faint targets is very sensitive to the variations in the background flux level. Even the best raw light curves we extracted contained additional variations up to $0.1-0.2$ mag in the last $10-20$~d of data. However, a simple high-pass frequency filter would remove any modulation signal along with the low-frequency noise. So instead we removed all low-frequency components below 0.5~d$^{-1}$ that were stronger than 10~mmag and were not connected to any potential modulation frequencies. The final light curves are displayed in the left panels of Figure \ref{fig:lcs}. The right panels show the folded light curves, along with the binned phase curves. 

The increasing background from the zodiacal light lowered the photometric accuracy of individual data points towards the end of the campaign. The accuracy of a single LC point was similar for EPIC~210282472 and 210282473: 0.050--0.055~mag at the beginning and 0.078--0.092~mag during the last days of the measurements. The errors are somewhat higher for the unblended star EPIC~210282474, starting from 0.087 mag and reaching 0.107 mag at the end of the campaign. 210282472 and 210282473 also appear to be brighter than 210282474 by 0.50 and 0.34~mag, respectively. As mentioned, the main component of this difference comes from the uncertainty of the zero-point determination for the confused sources. Similar differences appear in the peak-to-peak amplitudes that are lowered to 0.4 and 0.6~mag for 210282472 and 210282473, compared to 0.94~mag for 210282474. In contrast, \citet{moretti2009} found the amplitudes to be quite similar: 0.64, 0.73, and 0.65~mag in \textit{V} band, respectively. All these differences between EPIC~210282474 and the other two stars indicate that the confusing sources contribute considerably to the measured flux levels of EPIC~210282472 and 210282473 increasing their brightnesses and lowering the amplitudes and photon noise levels we measure. Nevertheless, this zero-point uncertainty only changes the overall amplitudes but does not distort the shape of the light curves, therefore has no further effect on the astrophysical content of the data.

For all of these three objects, photometric magnitudes have been 
transformed into USNO-B1.0 $R$ system \citep{monet2003}. 
This procedure has  been performed similarly as it was done by \cite{pal2015}.  
The raw and processed photometric data series are displayed in Table~\ref{table:phot}. 
Note that the photometric errors shown in this Table do not include 
the aforementioned zero-point uncertainty.

%% %% %% %% %% %% %% %% %% %% %% %% %% %% %% %% %% %% %% %% %% %% %% %% %% %% 
\begin{figure}
\includegraphics[width=1.0\columnwidth]{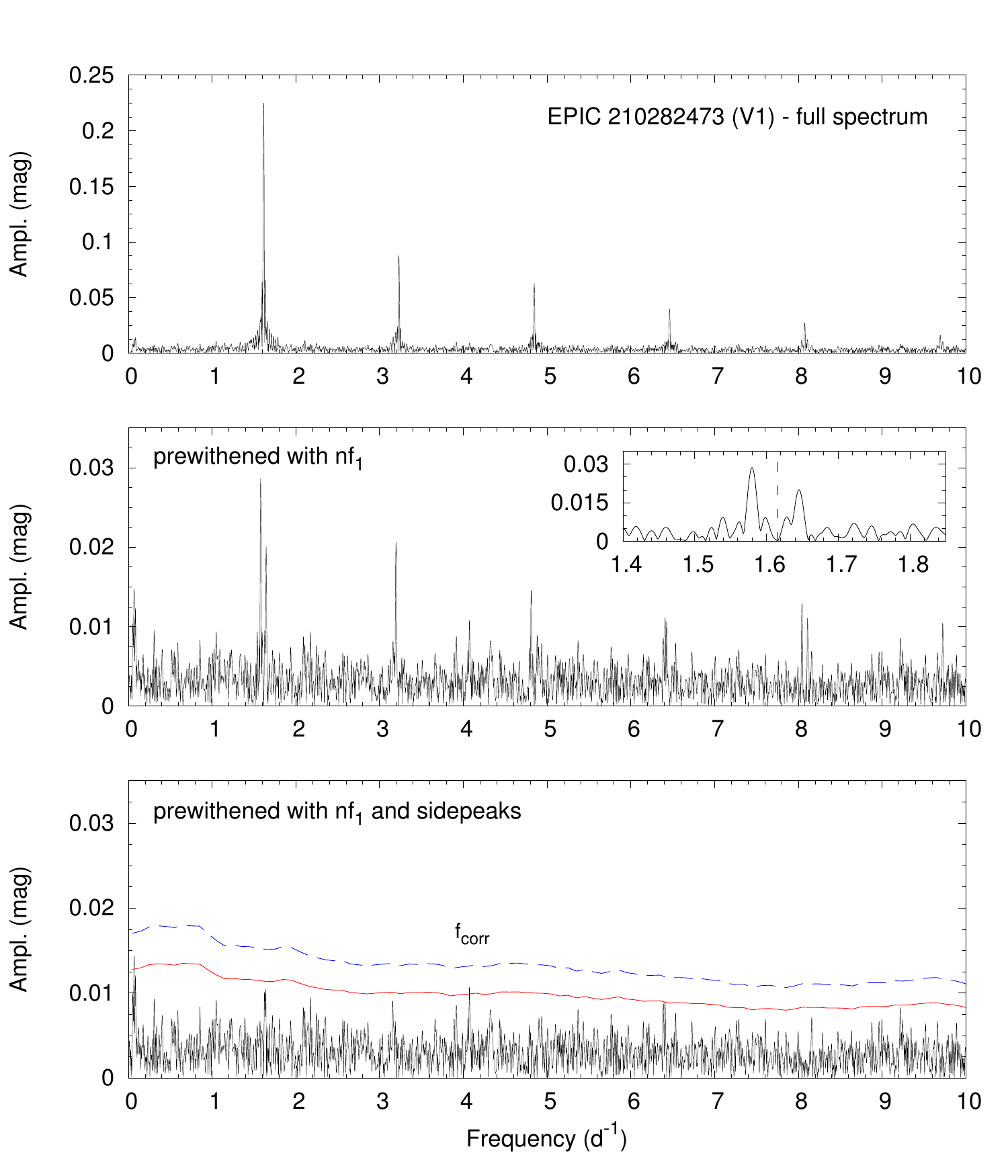}
\caption{Fourier transform of the light curve of EPIC 210282473. The top panel shows the original spectrum. Significant modulation sidepeaks are visible in the middle panel after prewhitening with the main peak and its harmonics. The insert shows the triplet around the position of the main peak (dashed line). The bottom panel shows the residual after the side peaks have been prewhitened as well. The red solid and blue dashed lines are the 3 and 4 SNR levels, respectively. The $f_{corr}$ label marks the position of the marginally detected spacecraft attitude correction frequency. }\label{fig:73_ft}
\end{figure}
%% %% %% %% %% %% %% %% %% %% %% %% %% %% %% %% %% %% %% %% %% %% %% %% %% %%

We carried out a standard Fourier analysis with the \textsc{Period04} software \citep{lenzbreger}, using multi-frequency least-squares fits and consecutive prewhitenings. The identified frequency components are listed in Table \ref{table:freq}. We detected modulation triplets ($nf_1$ harmonics and corresponding $nf_1 \pm f_m$ sidepeaks, where $n=1,2,3\dots$) in the Fourier spectra of EPIC~210282473 and 210282474. We note, however, that in the latter case the length of the data is insufficient to properly resolve the side-peaks, and their distances correspond to the length of the data instead of the modulation period, so we did not include them in the fit. 

We could also identify the characteristic signal of the spacecraft attitude correction maneuvers, at $f_{corr} = 4.08$~d$^{-1}$ and its harmonics, but with signal-to-noise ratios below 4. Apparently, very faint objects suffer less from the attitude changes of K2 than bright targets. Other significant frequency peaks, such as low-amplitude additional modes or half-integer peaks related to period doubling---both observed in many modulated RRab stars---did not appear after prewhitening. This is not surprising, however, considering that the 3~SNR (signal-to-noise ratio) significance limit in the Fourier spectra is around 8--12~mmag, well above the strongest additional peaks detected in the original \textit{Kepler} sample. The additional modes described by \citet{benko2014} reached only 3~mmag, so similar modes would be undetectable in the Leo IV stars. The Fourier spectra of EPIC~210282473 are shown in Figure \ref{fig:73_ft}.

Such faint targets are very sensitive to any instrumental problems arising both from the observations or the data reduction. Counterintuitively, the light curve of the only unblended star, EPIC~210282474 was the most problematic in our case. Even after the Fourier filtering, some slow, low-amplitude ($\sim0.1$ mag) variations still persisted in the last 15 days of the light curve (Figure \ref{fig:lcs}). Examination of the K2 full-frame images did not reveal any obvious contaminating sources in the same or adjacent modules. The SDSS images did not reveal any close-by objects either (see Figure \ref{fig:stamps}). We visually inspected the stacked image obtained using the observations taken with the 2.5~m Isaac Newton Telescope, the 4.2~m William Herschel Telescope, and the 4.1~m Southern Astrophysical Research Telescope (SOAR) in 2007 by \citet{moretti2009}, but found no significant nearby objects to within 1.0$^{\prime\prime}$. This image is displayed in Figure \ref{fig:leo4_map}. Images of the other two targets indicate that  the confusion is caused almost exclusively by a single bright galaxy near each star. Another star can be detected 2.1$^{\prime\prime}$ from 210282472, but it is about 3.5~mag fainter than the RR~Lyrae target so its contribution is negligible. 

%% %% %% %% %% %% %% %% %% %% %% %% %% %% %% %% %% %% %% %% %% %% %% %% %% %% 
\begin{deluxetable*}{llcccccc}
%\tabletypesize{\scriptsize}
\tablecolumns{7}
\tablewidth{0.95\textwidth}
\tablecaption{Frequency tables for EPIC
210282472, 210282473 and 210282474. \label{table:freq}}
\tablehead{EPIC & Freq.\ ID & \colhead{Freq (d$^{-1}$)} & \colhead{Amp (mag)} & \colhead{$\phi$ (rad/$2\pi$)} & \colhead{$\pm$F (d$^{-1}$)} & \colhead{$\pm$A (mag)} & \colhead{$\pm\phi$ (rad/$2\pi$)} }
\startdata
 & & & & & & \\
210282472 & $f_0$   &  1.40872  &  0.1400  &  0.1786  &  0.00011  &  0.0023  &  0.0024\\
-- &  $2f_0$  &  2.81745  &  0.0640  &  0.7512  &  0.00023  &  0.0023  &  0.0053\\
-- &  $3f_0$  &  4.22617  &  0.0427  &  0.380  &  0.00035  &  0.0023  &  0.008\\
-- &  $4f_0$  &  5.6349  &  0.0217  &  0.999  &  0.0007  &  0.0020  &  0.016\\
-- &  $5f_0$  &  7.0436  &  0.0106  &  0.679  &  0.0014  &  0.0019  &  0.032\\
\hline
 & & & & & & \\
210282473 & $f_0 $  &  1.61545  &  0.2249  &  0.1459  &  0.00008  &  0.0032  &  0.0017\\
-- & $2f_0$  &  3.23090  &  0.0869  &  0.6798  &  0.00020  &  0.0032  &  0.0045\\
-- & $3f_0$  &  4.84635  &  0.0616  &  0.2126  &  0.00028  &  0.0031  &  0.0064\\
-- & $4f_0$  &  6.46181  &  0.0405  &  0.801  &  0.00042  &  0.0031  &  0.010\\
-- & $5f_0$  &  8.0773  &  0.0265  &  0.349  &  0.0006  &  0.0030  &  0.015\\
-- & $6f_0$  &  9.6927  &  0.0152  &  0.950  &  0.0011  &  0.0028  &  0.026\\
-- & $f_0-f_m$  &  1.5818  &  0.0261  &  0.016  &  0.0006  &  0.0025  &  0.015\\
-- & $f_0+f_m$  &  1.6491  &  0.0150  &  0.344  &  0.0012  &  0.0031  &  0.029\\
-- & $2f_0-f_m$  &  3.1973  &  0.0199  &  0.485  &  0.0008  &  0.0030  &  0.019\\
-- & $3f_0-f_m$  &  4.8127  &  0.0132  &  0.998  &  0.0012  &  0.0030  &  0.028\\
-- & $4f_0-f_m$  &  6.4282  &  0.0104  &  0.507  &  0.0016  &  0.0031  &  0.037\\
-- & $5f_0-f_m$  &  8.0436  &  0.0119  &  0.144  &  0.0014  &  0.0031  &  0.033\\
-- & $5f_0+f_m$  &  8.1109  &  0.0098  &  0.436  &  0.0018  &  0.0029  &  0.041\\
-- & $6f_0+f_m$  &  9.7263  &  0.0095  &  0.097  &  0.0018  &  0.0029  &  0.043\\
\hline
 & & & & & & \\
210282474 & $f_0$ & 1.58582 & 0.2899 & 0.7674 & 0.00007 & 0.0032 & 0.0017\\
-- & $2f_0$ & 3.17164 & 0.1219 & 0.9159 & 0.00018 & 0.0032 & 0.0041\\
-- & $3f_0$ & 4.75747 & 0.0846 & 0.0845 & 0.00025 & 0.0036 & 0.0058\\
-- & $4f_0$ & 6.34329 & 0.0553 & 0.300 & 0.00039 & 0.0035 & 0.009\\
-- & $5f_0$ & 7.9291 & 0.0313 & 0.516 & 0.0007 & 0.0035 & 0.016\\
-- & $6f_0$ & 9.5149 & 0.0136 & 0.677 & 0.0016 & 0.0035 & ~0.036
\tablecomments{Frequencies, amplitudes, phases and their respective uncertainties. The modulation frequency for EPIC 210282473 was set to $f_m = 0.03361$ d$^{-1}$. }
\end{deluxetable*}
%% %% %% %% %% %% %% %% %% %% %% %% %% %% %% %% %% %% %% %% %% %% %% %% %% %% 
%% %% %% %% %% %% %% %% %% %% %% %% %% %% %% %% %% %% %% %% %% %% %% %% %% %% 
\begin{figure}
\includegraphics[width=1.0\columnwidth]{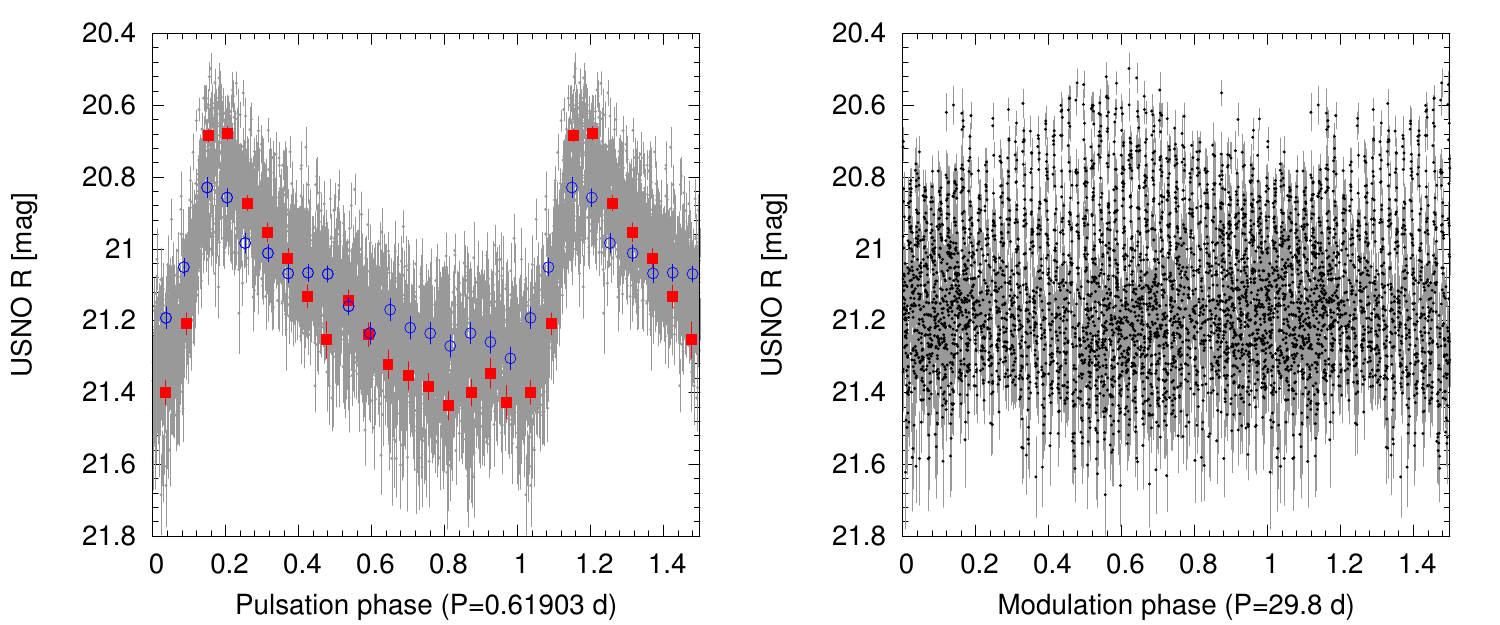}
\caption{Blazhko effect in EPIC~210282473. Left: data folded with the pulsation period. Binned data from 2-day long 
sections are overlaid, red dots show the maximum-amplitude modulation phase, blue circles show the minimum-amplitude
phase.  Right: light curve folded with the modulation period. }\label{fig:73_mod}
\end{figure}
%% %% %% %% %% %% %% %% %% %% %% %% %% %% %% %% %% %% %% %% %% %% %% %% %% %%

%% %% %% %% %% %% %% %% %% %% %% %% %% %% %% %% %% %% %% %% %% %% %% %% %% %% 
\begin{figure*}
\includegraphics[width=1.0\textwidth]{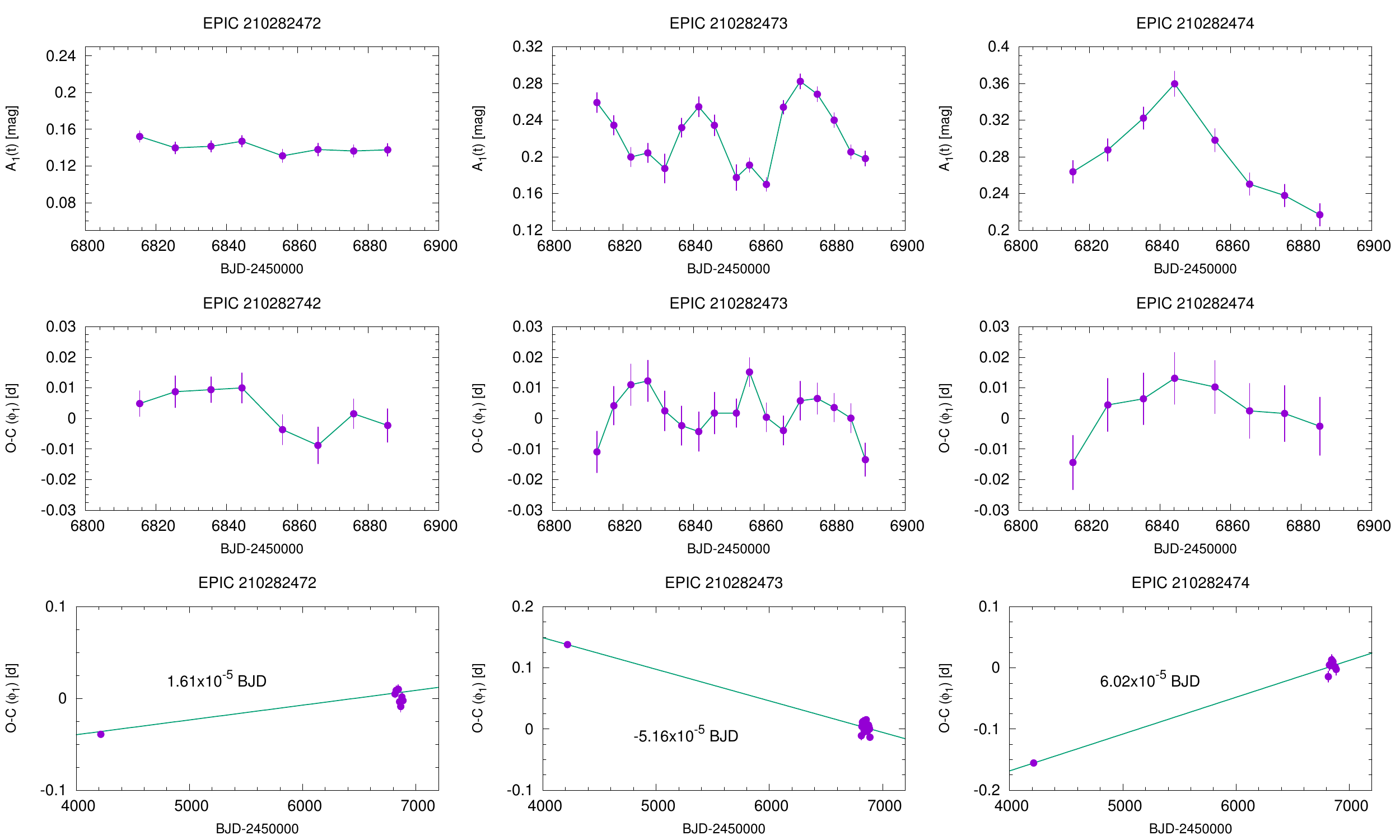}
\caption{Top row: variation $A_1$, the amplitude of the $f_1$ frequency component in all three stars. Middle row: O--C variation, calculated as the variation of the $\phi_1$ Fourier-parameter. Bottom row: long-term period drifts of all three stars between 2007 \citep{moretti2009} and 2014, the K2 observations. }\label{fig:leo4_oc}
\end{figure*}
%% %% %% %% %% %% %% %% %% %% %% %% %% %% %% %% %% %% %% %% %% %% %% %% %% %%

\section{Results}
\label{sec:results}
\subsection{Blazhko effect in Leo IV}
One star, EPIC~210282473, is clearly modulated. Although the signs of amplitude changes in RR~Lyrae stars have been observed as far as the Andromeda galaxy \citep{brown2004}, this is the first star beyond the Magellanic Clouds for which the detailed modulation properties can be determined. The K2 data covers almost three Blazhko cycles. The distance of the triplet side peaks ($nf_1\pm f_m$) suggested a modulation period of $P_m = 30\pm 3$~d, but the independent fits to the modulation peaks resulted in large uncertainties. We also calculated the amplitude and phase variations ($A_1$, $\phi_1$) of the main frequency peak ($f_1$) by dividing the data into short segments. A sine fit to $A_1(t)$ itself led to a more precise Blazhko period: $P_m = 29.8\pm0.9$~d. 

Figure \ref{fig:73_mod} shows the modulation properties of EPIC~210282473. The right panel is the light curve, folded with the modulation period: the presence of the Blazhko effect is even more clear here than in the normal light curve. The left panel illustrates how the shape of the light curve itself changes over a modulation cycle. We selected two 2-day long sections of a low- and a high-amplitude state, respectively, and calculated the phase-binned values of the data. Red squares and blue circles show the two states, respectively: the differences in amplitude are evident around maximum and minimum light, but the phases at which the extrema occur show little to no shifts.

Another representation is included in Figure \ref{fig:leo4_oc} where the upper and middle panels show the variation of the $A_1$ and $\phi_1$ Fourier-terms. These figures confirm that there is significant amplitude variation in the data, but the phase modulation is almost negligible in the star. Moreover, the middle column of Figure \ref{fig:leo4_oc} indicates that the cycles are not repetitive and the amplitude and phase variation are not strictly correlated to each other, similar to some modulated stars in the original \textit{Kepler} sample \citep{benko2014}.

The case of EPIC~210282474 is more ambiguous, since the light curve of the star suffers from instrumental effects more than those of the other two. The Fourier-filtered light curve shows some amplitude and phase variation but those are suspiciously---but not exactly---symmetric to the mid-campaign data download period. This symmetry in the amplitudes invites to suspect some kind of instrumental origin, although we could not identify any contaminating source. We also detect weak phase variations that seem to follow the changes in pulsation amplitude (right column of Figure \ref{fig:leo4_oc}). If we accept that these variations indeed originate from modulation, its period is clearly longer than the length of the campaign. We identified side-peaks in the frequency spectrum, but given the limited length of the data, they are not resolved properly. Overall, since we detect the variations in the first, better-quality part of the light curve as well, we accept this signal as a likely (although not unquestionable) detection of the Blazhko effect. Most stars in the Milky Way have modulation periods longer than 10--20~d \citep{szczfab2007}: the values we determined for the Leo IV stars agree with that period distribution. Although the Leo IV sample is very limited, it confirms that the Blazhko effect is abundant in other galaxies as well. 

EPIC~210282472 also shows some phase variations but only marginal changes in amplitude (left column of Figure \ref{fig:leo4_oc}). However, these amplitude changes can be partly attributed to the large scatter of the light curve points that make the phase determination uncertain for shorter data segments (10 days per segment in this case). Given the lack of significant amplitude variations, we classify this star as non-modulated.

We also calculated the Fourier phase values for the original observations of \citet{moretti2009} with the frequencies determined from the K2 data, and transformed them to the more classical O--C diagrams. The results, shown in the bottom row of Figure \ref{fig:leo4_oc}, suggest that pulsation period of EPIC~210282473 got shorter during the 7.1~yr separating the two data sets while the periods of EPIC~210282472 and 210282474 got longer. Other observations between the two epoch were too sparse to determine the pulsation phase and O--C values. Therefore actual rates of period change cannot be determined yet.

%% %% %% %% %% %% %% %% %% %% %% %% %% %% %% %% %% %% %% %% %% %% %% %% %% %% 
\begin{deluxetable}{lccc}
%\tabletypesize{\scriptsize}
\tablecolumns{4}
\tablewidth{0pc}
\tablecaption{Pulsation periods, modulation periods and photometric [Fe/H] indices of the RR Lyr targets. \label{table:stellarparam}}
\tablehead{Object & Pulsation per.\ (d) & Mod.\ per.\ (d) & [Fe/H]}
\startdata
472 & $0.709862\pm 5.5\cdot10^{-5} $ & --- & $-2.16\pm0.10$ \\
473 & $0.619022 \pm3.1\cdot10^{-5} $ & $29.8\pm0.9$ & $-2.64\pm0.10$ \\
474 & $0.630588 \pm4.4\cdot10^{-5} $ & $>80$ & $-2.62 \pm0.15$
\enddata
\end{deluxetable}
%% %% %% %% %% %% %% %% %% %% %% %% %% %% %% %% %% %% %% %% %% %% %% %% %% %% 

\subsection{Photometric metallicities}
A great advantage of RR~Lyrae stars is that we can estimate their [Fe/H] indices from photometry alone \citep{jurcsikkovacs96}. \citet{nemec2013} have calibrated the relation between the light curve parameters and the spectroscopically determined metallicity values for the \textit{Kepler} passband and showed that the photometric and spectroscopic [Fe/H] indices agree to $\pm0.1$~dex for all but the extremely modulated stars. We calculated the photometric [Fe/H] values for all three stars with this method: the results are summarized in Table \ref{table:stellarparam}. The accuracy intrinsic to Fourier parameter determination is $\pm 0.03$~dex. However, according to \citet{nemec2013}, the uncertainties of the relation are dominated by the calibration of the equation and the determination of the spectroscopic values, so we assumed higher uncertainties for all stars (Table \ref{table:stellarparam}.  

Of the three stars, only EPIC~210282472 has a spectroscopically determined index, ${\rm [Fe/H]} = -2.03$ \citep{kirby2008} that is higher, but within the respective error bars compared to our photometric index of $-2.16\pm0.10$, assuming the same uncertainties. \citet{moretti2009} determined the photometric metallicity for EPIC 210282473, but their result, ${\rm [Fe/H]} =-2.11$ is significantly higher than the K2-derived value ($-2.64\pm0.10$). This discrepancy can largely be attributed to the fact that the linear relation of \citet{jurcsikkovacs96}, used by \citet{moretti2009}, overestimates the [Fe/H] index by as much as $\sim 0.3$ dex for very metal-poor stars. The K2 data also revealed that the star is modulated, and the poor modulation phase coverage may have affected the original metallicity determination. Since the Blazhko effect in the star is well covered by the K2 data and does not show extreme amplitude changes or cycle-to-cycle variability, we consider the photometric metallicity accurate to 0.10 dex  \citep{nemec2013}.

We calculated the metallicity of the third star, EPIC~210282474, to be ${\rm [Fe/H]} = -2.62\pm0.15$. We note that if the star is indeed modulated, the partial phase coverage of the Blazhko effect could affect the photometric metallicity value. We increased the uncertainty to account for this additional factor. Overall, the three metallicity values are in good agreement with the average metallicity values of Leo IV (around --2.3 or --2.6) and confirm its very metal-poor nature. 

\newpage

\section{Conclusions}
\label{sec:conclusions}
We detected the variations of all three extragalactic RR~Lyrae stars in the K2 mission Campaign~1 data. The stars are members of the ultra-faint dwarf spheroidal galaxy Leo~IV \citep{moretti2009}. These are the faintest pulsating stars measured with the \textit{Kepler} space telescope so far, at a brightness level of $Kp \approx 21.5$ mag. 

One of the variables, EPIC~210282473, displays clear amplitude variations with a period of $29.8\pm0.9$ days, while EPIC~210282474 is likely modulated with a period longer than the campaign length (80~d). In the case of the former star, almost three modulation cycles are covered, during which the strength of the Blazhko effect and the relation between the amplitude and phase modulation both changed. These observations are the farthest measurements of the detailed parameters of the Blazkho effect at a distance of 154~kpc, and the first one beyond the Milky Way and the Magellanic Clouds. We determined the photometric [Fe/H] indices of the stars that range between --2.16 and --2.64, in agreement with the spectroscopic values measured in Leo~IV.

We also demonstrated that with fine-tuned algorithms based on image subtraction, nearly the same faintness level can be achieved for stationary targets that is attainable for moving objects, such as TNOs. We found that unlike brighter sources, very faint but high amplitude variations do not significantly contain frequencies caused by the attitude correction maneuvers of \textit{Kepler}. The implications of this exercise are far-reaching, as photometric space missions will continue to stare at crowded fields in the future. Globular clusters have already been measured during the K2 mission, and \textit{Kepler} may observe Cepheid stars in another galaxy, IC~1613, in Campaign 8. The TESS and PLATO missions will employ even larger pixels that \textit{Kepler} did, leading to strong crowding in several fields where image subtraction will be a necessary requirement \citep{tess,plato}. Just one, important example is the Large Magellanic Cloud within the southern continuous observing zone of TESS. The LMC contains several bright, large-amplitude variables (Cepheids for example) that will be accessible for that mission, but their photometry will present challenges very similar to the K2 observations presented here in terms of crowding and signal-to-noise ratios. 

%\vspace{-5mm}

\begin{acknowledgements}
We are grateful to the referee for the constructive comments that helped to improve the paper. We thank the hospitality of the Veszpr\'em Regional Centre of the 
Hungarian Academy of Sciences (MTA VEAB) where part of our work was carried out. 
Funding for the \textit{Kepler} and K2 missions is provided by the NASA Science 
Mission directorate. The authors gratefully acknowledge the Kepler team,
the Guest Observer Office, and Ball Aerospace, whose outstanding efforts have made these results possible. 
This project has been supported by the Lend\"ulet-2009, 
LP2012-31 and LP2014-17 Programs of the Hungarian Academy of Sciences
and the OTKA grants K-104607, K-109276, and K-115709.
The research leading to these results has received funding from the ESA PECS Contract No.\ 4000110889/14/NL/NDe,
 and from the European Community's Seventh Framework Programme (FP7/2007-2013) 
 under grant agreements no.\ 269194 (IRSES/ASK) and no.\ 312844 (SPACEINN). 
 L.M.\ was supported by the J\'anos Bolyai Research Scholarship of the Hungarian Academy of Sciences. 
 VR acknowledges partial support by the  project PRIN-MIUR (2010LY5N2T) 
``Chemical and dynamical evolution of the Milky Way and Local Group galaxies" (P.I.: F.\ Matteucci). 
All of the data presented in this paper were obtained 
from the Mikulski Archive for Space Telescopes (MAST). 
STScI is operated by the Association of Universities for Research in Astronomy, Inc., under 
NASA contract NAS5-26555. Support for MAST for non-HST data is provided by the 
NASA Office of Space Science via grant NNX09AF08G and by other grants and contracts.
\end{acknowledgements}

%\bibliographystyle{aj}
%\bibliography{K2asteroids}

{}

\end{document}